# Label-free identification of biomolecules by single-defect-spectroscopy at the aqueous hexagonal boron nitride interface

Miao Zhang*, Cristina Izquierdo Lozano, Stijn van Veen, and Lorenzo Albertazzi*

Department of Biomedical Engineering, and Institute for Complex Molecular Systems, Eindhoven University of Technology, 5600MBEindhoven, The Netherlands

m.zhang1@tue.nl, l.albertazzi@tue.nl

**ABSTRACT**

Label-free single-molecule detection is essential for studying biomolecules in their native state, yet having materials with intrinsic molecular specificity to identify a broad range of molecules without complex functionalization remains challenging. We present a method that utilizes emissions from selectively activated defects at the aqueous hexagonal boron nitride (hBN) interface to detect and identify biomolecules, including lipids, nucleotides, and amino acids. Using spectrally-resolved single-molecule localization microscopy combined with machine learning, we harvest spatial, spectral and temporal data of individual events to uncover optical fingerprints of biomolecules. This approach allows us to probe fine chemical differences, as small as the single deprotonation of an amino acid's side chain and detect dynamics of biomolecules at the interface with exceptional detail. As a proof-of-concept, we identified five different amino acids at the single-molecule level with high accuracy. Our findings shed light on hBN-biomolecule interactions and highlight the potential of hBN for label-free single-molecule identification.

**INTRODUCTION**

Single-molecule studies are crucial for understanding molecular dynamics and diversity on the microscopic scale[1–5]. Especially, label-free approaches that preserve molecules in their native states are particularly desirable[4,6–8]. However, nearly all optical label-free methods rely on functionalizing the sensing materials with target-specific binders, typically antibodies, for molecular identification. This process is often cumbersome and chemically complex, limiting the sensor's versatility, reproducibility, and ability to detect multiple molecular species within the same matrix[9]. The need for materials capable of label-free detection with intrinsic molecular specificity remains unmet. Such materials should ideally exhibit a distinct response to fine chemical differences without requiring external modifications.

Hexagonal boron nitride (hBN) offers significant potential as a label-free detection platform due to its multifarious emissive defects in atomically smooth surface[10–12] that are sensitive to the environmental changes. Recent studies have shown that defect states of hBN within its 6 eV bandgap can be activated by various organic solvents to emit photons with distinct intensities and wavelengths[13,14]. However, using hBN for molecular specific analysis of complex biomolecules in aqueous solutions has so far never been demonstrated, partially due to the strong quenching effect of water[13,14].

In this work, we present a strategy for label-free single-molecule detection and identification by exploiting the specific interactions between biomolecules and surface defects in hBN. These transient interactions result in sparse, intermittent photon emissions that carry molecular-specific information. We exploit spectrally-resolved Point Accumulation for Imaging in Nanoscale Topography (sPAINT) microscopy to capture this rich information. The PAINT modality works by localizing individual emitting events over time to build a highly detailed map of molecular interactions[15]. By adding a dispersive optical component, sPAINT also captures the individual spectrum of each emitting event[16–18]. Through sPAINT, we visualize these transient interactions on a large scale and obtain spatial, spectral and temporal information of the emissions simultaneously. This multidimensional dataset is then processed using machine learning algorithms to uncover molecular fingerprints for identification of biomolecules.

With this approach, we explored a wide range of fundamental small molecular components of life, including lipids, amino acids, and nucleotides. We showed that biomolecules that strongly bind to the hBN surface in aqueous solutions can activate defect emissions from both pristine and plasma-treated hBN in PBS buffer. We found that different chemical structures of biomolecules selectively activate different populations of defects with distinguishable emission wavelengths and binding kinetics. Leveraging on this principle, we reveal single-molecule interactions with the hBN surface with unprecedented details. Building on top of this discovery, we use hBN as a sensor for biomolecules and based on the spatial and spectral signature of molecule-activated defect emissions, we could identify five different amino acids at the single-molecule level by exploiting a machine learning-assisted classification algorithm. Our findings unravel the complex interactions of hBN with biomolecules and demonstrate it as a promising material for label-free biomolecule detection and identification.

**Molecule-activated emissions from defects in hBN in aqueous solution**

To investigate the interaction between molecules and the hBN defects in aqueous solution, we utilized sPAINT microscopy to capture spatial and spectral images simultaneously. This was achieved by placing a transmission grating in front of a camera attached to a total internal reflection fluorescent microscope (Fig. 1a and Methods). Using a 532 nm laser for excitation, we monitored emissions on hBN surface in the range of 540-750 nm at a frame rate of 20 Hz. Thin hBN flakes, exfoliated from commercially available high-quality crystals and transferred onto a glass coverslip, were treated with air plasma to introduce defects (Methods). Remarkably, strong photoluminescence (PL) blinking was observed on the hBN surface when in contact with a wide range of biomolecules in buffer solutions. A typical PL image of the phospholipid DOPC on the hBN surface is shown in Fig. 1b, demonstrating great signal-to-noise ratio. We attribute these PL signals to defect emissions in hBN surface, which are activated by the transient molecular binding.

To get an insight into molecule-activated defect emissions and to find fingerprints for each molecular event, we conducted a quantitative analysis of spatial, spectral and temporal data of emitting events, followed by machine-learning based molecule identification (Fig. 1c). From the spatial images (Fig. 1a), intensities and locations of emissions were mapped with high precision down to 20 nm by single-molecule localization algorithm (Methods). The corresponding single-defect spectra were extracted from the spectral images (Fig. 1a) simultaneously with identical time stamps. Our custom analysis script then calculated key parameters from the extracted data: i) event density, reflecting molecule affinity and defect activation; ii) features of single spectrum and histogram of emission wavelengths, including spectral intensity, width, mean wavelength, peak wavelength, skewness, and kurtosis, depicting the activated defect populations; iii) event duration and diffusion coefficient of molecule diffusion, indicating binding kinetics between molecule and defect (Methods). The multidimensional data was then used for training Random Forest models to recognize complex patterns for identifying molecules that otherwise cannot be distinguished by single or paired parameters.

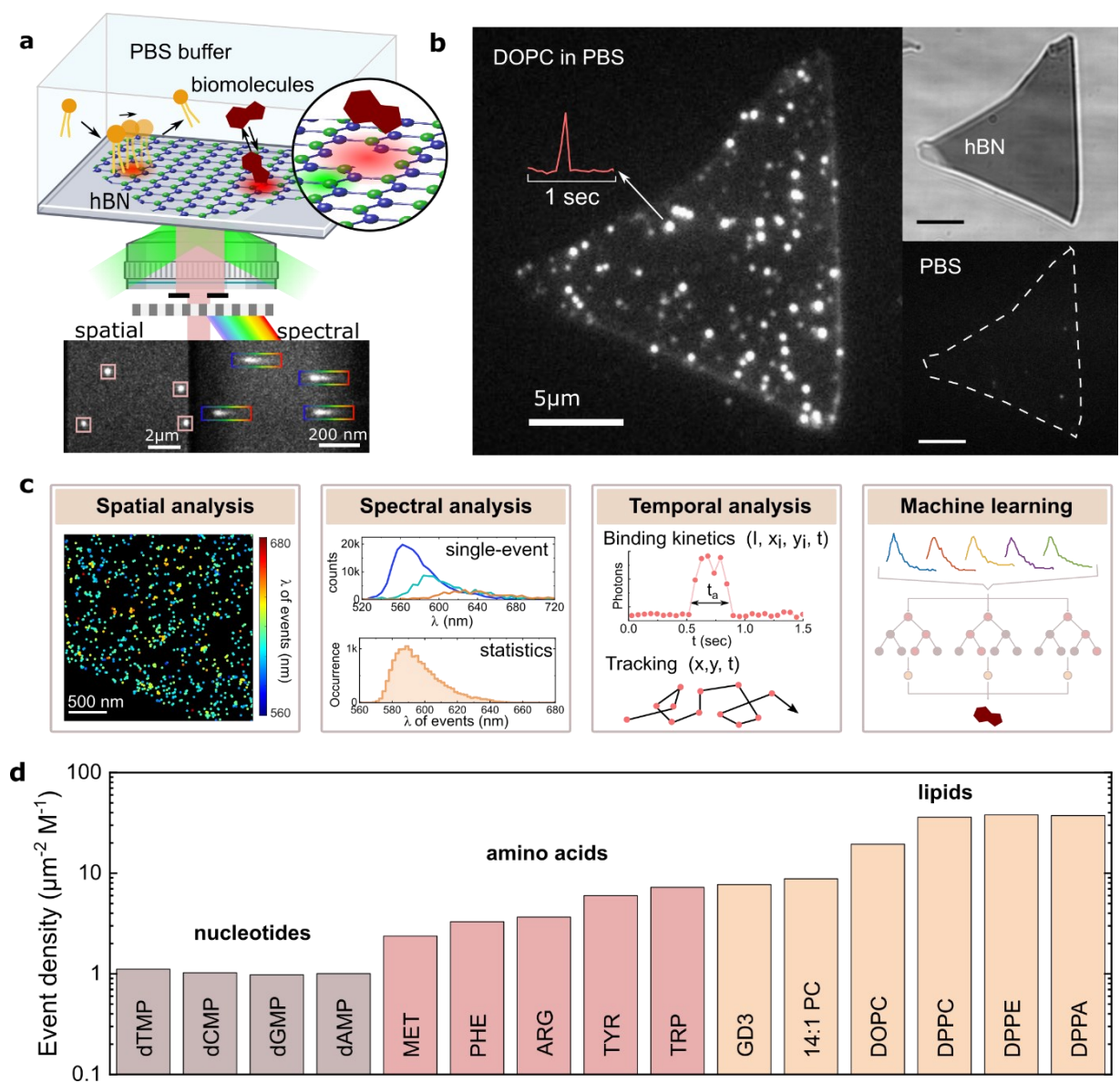

**Figure 1. Molecule-activated emissions from surface defects in hBN in aqueous solution. a**, Schematic of spectrally-resolved PAINT microscopy to detect molecule-activated emissions from surface defects in hBN. The spatial position of the emission and the corresponding spectrum were obtained simultaneously from the 0th-order and 1st-order diffraction of a transmission grating that is placed before the camera. **b**, Photoluminescence images of a hBN flake in PBS buffer with or without DOPC lipid molecules and a white light image of the hBN flake. The concentration of DOPC is 550μM. . **c**, Typical analysis of the molecule-activated emission consists of spatial analysis, spectral analysis of single events and statistics of emission wavelengths on a hBN flake, and temporal analysis of molecule/defect binding kinetics and tracking of molecule diffusion on the hBN surface. ISingle molecules were identified based on their spatial and spectral features through the use of machine learning algorithms. **d**, Normalized event density ($\mu m^{-2} M^{-1}$) of molecule-activated defect emissions on hBN with nucleotides, amino acids and lipid molecules in PBS, in semi-log scale.

We explored the defect emissions activated by a wide range of biomolecules, covering the major classes of small biomolecules: i) phospholipids and glycolipids (DPPE, DPPA, DPPC, DOPC, 14:1 PC, ganglioside GD3), ii) amino acids (Trp, Tyr, Arg, His, Met, Phe, Pro, Val, Lys, Gly) and iii) nucleotides (dTMP, dCMP, dGMP, dAMP). Remarkably, all molecules activated defect emissions. To compare the molecular signal, we normalized event density per frame by the concentration of analyte in solution (Fig.

1d). Phospholipids with long hydrocarbon chains exhibited the highest event density (> 10 $\mu m^{-2}$ $M^{-1}$), while amino acids induced event density in the range of 1-10 $\mu m^{-2}$ $M^{-1}$, generally correlating to their side-chain sizes. Nucleotides showed the lowest event density (~1 $\mu m^{-2}$ $M^{-1}$). Emissions in PBS buffer resulted in background event densities, typically in the order of $10^{-4}$ $\mu m^{-2}$. The concentration of analyte was adjusted to achieve event density at least one order of magnitude higher than the background (Fig. S1). This ability to detect biomolecules is remarkable, given that water was found as an effective quencher of defect emissions, as 30% water in acetonitrile is enough to quench most of the defect emissions activated by acetonitrile (Supplementary discussion and Fig. S2), consistent with the previous report[14]. Our results suggest that molecules strongly adsorb to the hBN surface in aqueous solutions, whichcan block water molecules, thereby activating PL emissions of defects. The interaction likely occurs through the hydrophobic or planar moieties of the molecules, as larger hydrophobic or planar moieties correlate with higher event densities. These insights offer a powerful platform for studying hBN-biomolecule interactions and for detecting biomolecules using hBN. In the following sections, we will discuss the detection of different classes of molecules in detail.

**Interaction between lipids and hBN surface**

The amphiphilic lipid molecules triggered defect emissions effectively. To investigate the interactions of lipids on the hBN surface and to find out the chemical moieties that are responsible for defect activation, we examined the defect emissions activated by lipids with different head groups (DPPC, DPPE, DPPA, ganglioside GD3), and with hydrocarbon chains differing in length and number of unsaturated bonds (DPPC, DOPC, 14:1 PC) (Fig. 2a). Lipids were prepared as multilamellar vesicles in PBS buffer (Methods). The wavelengths of defect emissions show clear differences between different types of lipid molecules, as can be seen in Fig. 2b,c. Phospholipids consistently activated defects emitting at ~590 nm, whereas the glycosphingolipid ganglioside GD3 activated a distinct population of defects emitting around 620 nm. A difference was also observed in their event durations with phospholipids showing a longer bound time (Fig. S3). We attribute this to the selective activation of different defect structures by specific chemical groups. The phospholipids likely adsorb on the hBN surface through hydrophobic interaction with their hydrocarbon chains, which has been previously observed on graphene surfaces[19,20]. The hydrocarbon chains can activate a population of defects due to their simple structure, regardless of chain length or number of unsaturated bonds. This aligns with a previous report identifying carbon as a crucial element in defect emissions around 585 nm in hBN[12]. In contrast, emissions activated by ganglioside GD3 at 620 nm are likely caused by the glycan head group. Moreover, varying defect populations by surface treatment of hBN also led to different emission histograms, with GD3-activated defects on pristine hBN showing a redshift compared to air plasma treated hBN (Fig. 2d). The difference is also reflected in their corresponding event durations, with pristine hBN showing a population of longer bound events (Fig. S4). Similar behavior was also observed in acetonitrile-activated events on different hBN surfaces (Fig. S2e). These results clearly demonstrate the specificity of the molecule-defect interactions, which is crucial for molecule identification.

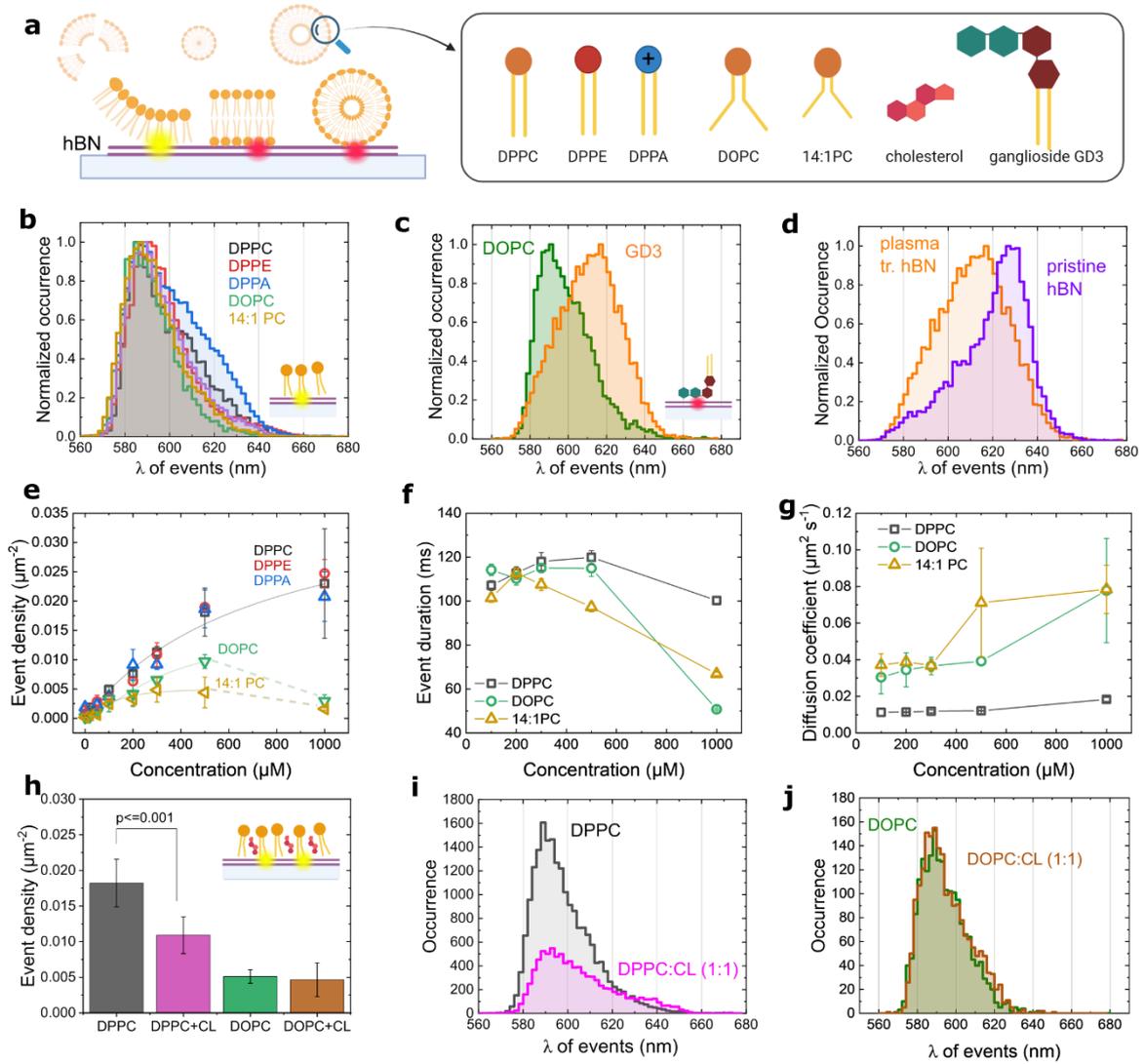

**Figure 2. Interactions between lipids and hBN surface in PBS. a**, Schematic of interactions between various lipids and the hBN surface. **b**, Normalized histograms of wavelengths of defect emissions activated by DPPC, DPPE, DPPA, DOPC, 14:1 PC at 500 µM. **c**, Normalized histograms of wavelengths of defects emission activated by DOPC and ganglioside GD3 at 500 µM. **d**, Normalized histogram of event wavelengths on plasma treated hBN and pristine hBN activated by GD3. **e**, Adsorption curves of various lipids. The adsorption isotherms were described by a general surfactant adsorption isotherm equation[21,22] (details in SI). **f**, Event duration of emissions activated by DPPC, DOPC and 14:1 PC. The times were extracted by fitting double-exponential decay functions to the histogram of event durations collected from 3 different hBN flakes at each concentration. Error bar is standard error. **g**, Diffusion coefficients of DPPC, DOPC and 14:1 PC, calculated from the distribution of instantaneous velocities on 3 different hBN flakes at each concentration. Error bar is standard deviation. **h**, Event density of DPPC, DOPC and their mixture with cholesterol (1:1). **i**, histogram of event wavelengths of DPPC and a mixture of DPPC and cholesterol (1:1). **j**, histogram of event wavelengths of DOPC and a mixture of DOPC and cholesterol (1:1). Concentrations in h, i, j are 500 µM.

Further, we used lipid-activated defect emissions to monitor kinetic and dynamic changes of lipid interactions with hBN at various lipid concentrations. As can be seen in Fig 2e, phospholipids with different head groups, but identical hydrocarbon chains (DPPC, DPPE, DPPA) show similar adsorption curves, yet altering the length and the number of unsaturated bonds of hydrocarbon chains significantly affects event densities of DOPC (18:1 PC) and 14:1PC. These again indicate that phospholipids interact with hBN surface via hydrocarbon chains. The event density induced by DPPC started to saturate above 1 mM, while DOPC and 14:1PC peaked around 500 μM and 300 μM, respectively, and then decreased. As event density saturated, a significant decrease of event duration was observed for DOPC and 14:1PC (Fig. 2f), alongside increased lateral diffusion coefficients (Fig. 2g), suggesting altered binding kinetics and dynamics at higher concentrations. Such observations indicate that at higher concentrations, the interaction between lipid molecules affect the binding between lipid and defect, lowering the energy barrier for lipid hopping between defects. Fast lipid diffusion reduces the binding time to defects, reducing the detectable events, as observed for DOPC and 14:1PC. No significant change in event duration and diffusion coefficient were observed for DPPC. These agree well with the literature that at room temperature DOPC in a lipid membrane has a lateral diffusion coefficient two orders of magnitude higher than DPPC[23–25]. However, our ability to observe the fast dynamics at high lipid concentrations is limited by our temporal resolution (20 fps). Additional tracking measurement at a faster frame rate (48 fps) and 1mM sample concentrations confirmed the diffusion coefficients in the same range as reported for lipid monolayers in the literature[23–25] (Supplementary discussion and Fig. S6). In addition, staining lipids with lipophilic dyes showed full coverage of lipids on the hBN surface (Supplementary discussion and Fig. S7). We speculate that a continuous monolayer of lipids was formed on the hBN surface at high concentrations.

We further investigated how lipid packing influences defect activation by adding cholesterol to DPPC and DOPC to induce phase changes. As shown in Fig. 2h, adding cholesterol to DPPC (1:1 ratio) significantly reduced the event density ($P < 0.001$), while adding cholesterol to DOPC (1:1 ratio) showed negligible change, with both showing much lower event densities compared to DPPC. The wavelength histograms revealed that the main population of emissions remained around 590 nm for all samples (Fig. 2i, j), implying that the defect activation occurs via hydrocarbon chains regardless of cholesterol presence. These variations in defect activation align well with the changes in lipid phases induced by cholesterol. Cholesterol addition shifts the DPPC membrane from a solid-gel phase (Lβ) or liquid-condensed phase (LC) to a liquid-ordered phase (Lo), characterized by less ordered hydrocarbon chain packing[26–29]. In contrast, the DOPC membrane remains in liquid-disordered phase (Ld) or liquid-expanded phase (LE)[27–29], with highly disordered packing[30–32]. Our results suggest that the defect activation rate is positively correlated to the lipid packing order, where more confined hydrocarbon chains and a lower water permeability in a more orderly lipid patches[33] increase defect activation rates (Supplementary discussion). These findings highlight the sensitivity of defect activation to lipid conformational changes, establishing this platform as a valuable tool for studying membrane fluidity, permeability, and phase transition, which are relevant to their functionality.

**Defects emission activated by amino acids**

Amino acids are ideal molecules to investigate the ability of activated defect emissions to discriminate biomolecules, as they can vary significantly in side chain structure. As depicted in Fig. 3a, ten amino acids were chosen, tryptophan (Trp), tyrosine (Tyr), phenylalanine (Phe), histidine (His), arginine (Arg), lysine (Lys), valine (Val), methionine (Met), proline (Pro) and glycine (Gly). Remarkably, all tested amino acids induced distinct defect emissions with different wavelengths and densities (Fig. 3b and c). This suggests that the side chains of amino acids mediate their interaction with hBN, consistent with previous molecular dynamic studies on amino acids adsorption on 2D materials[34–37]. Complex physiochemical properties of the side chains, such as surface area, hydrophobicity, and charge state, can all affect the activation of defect emissions, resulting in varying wavelengths and densities.

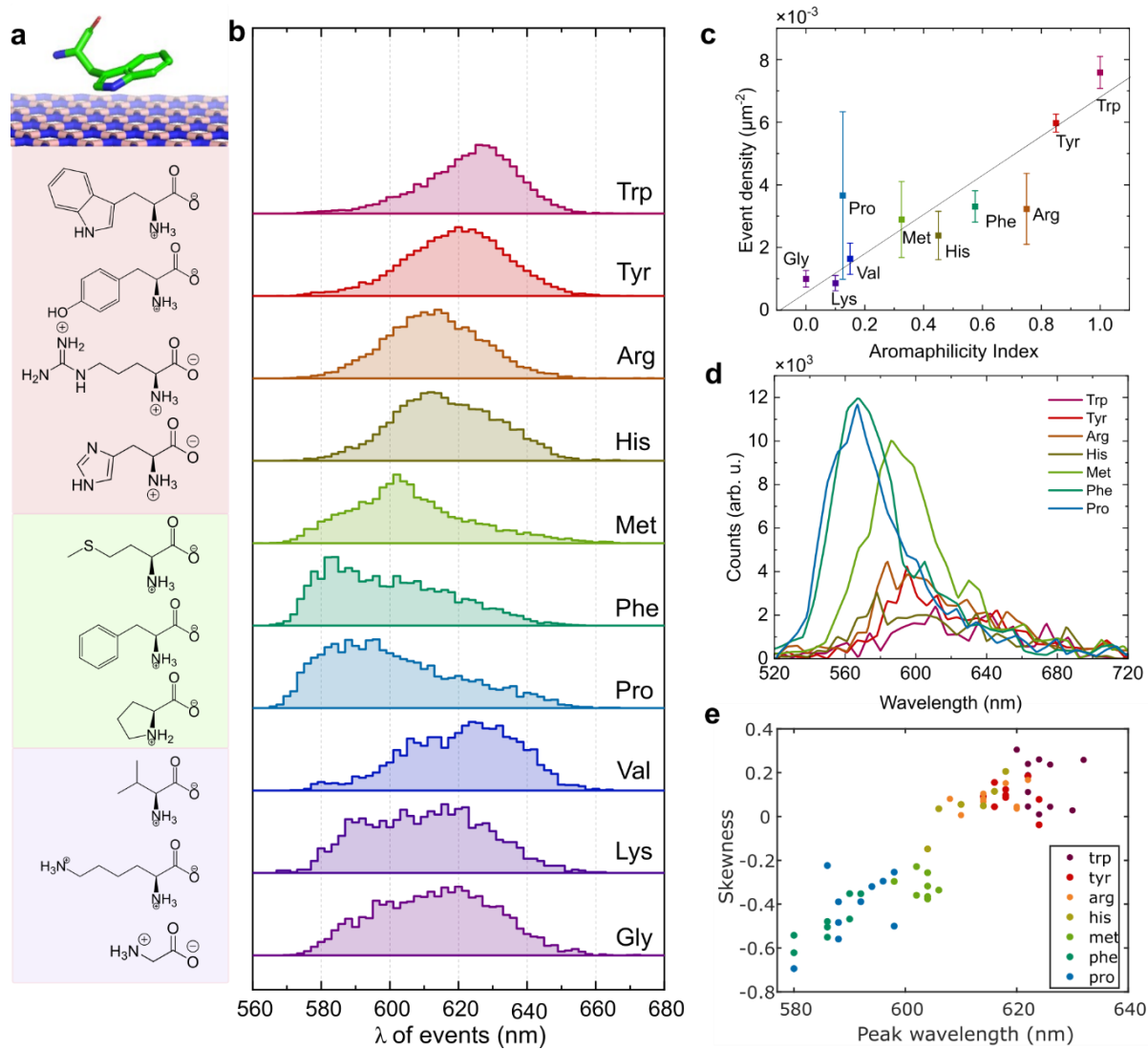

**Figure 3. Defects emission activated by amino acids show chemical sensitivity. a**, A schematic illustration of the interaction between amino acids and hBN surface via side chains. 10 amino acids were chosen in this work. **b**, Normalized histograms of defects emission wavelength activated by 10 chosen amino acids. **c**, Event density of amino acids versus aromaphilicity index. Concentration: 1 mM **d**, Single defect spectrum of most probable events activated by different amino acids. **e**, A scatter plot of peak wavelength of emission histogram versus skewness of the histogram. Each dot represents ensemble defect emission features extracted on a 5-by-5 µm hBN surface over 22K frames.

Normalized histograms of emission wavelengths of defects show that the population of emitting defects shifts when interacting with different amino acids (Fig. 3b), indicating a selective activation of defects by the various amino acids. Amino acids with large planar side chains (Trp, Tyr, Arg, His) predominantly activated defects emitting between 610 – 630 nm, while those with smaller, weakly electron-donating side chains (Met, Phe, Pro) [38] activated defects emitting between 580 – 600 nm. Spectra of single events show that emissions below 600 nm were generally brighter than emissions in the longer wavelength range (Fig. 3d). Interestingly, although Phe and His both have planar side chains with similar surface areas[39], they activate distinct defect populations, likely due to differences in their electronic properties (Table 2 in SI). His is slightly positively charged as an electron acceptor at neutral pH, whereas Phe is a weak electron donor [38]. This implies that defects emitting between 580 - 600 nm may act as electron acceptors, activated by electron donating structures below a certain size. This is consistent with electron donating hydrocarbon chain-activated defect emissions in the same wavelength range as we discussed for phospholipids previously. On the other hand, defects emitting between 610 - 630 nm appear to be activated primarily by large planar structures, regardless of electronic properties, indicating the distinguishing feature is the planar size of the chemical structure. This aligns with the defect emissions observed with ganglioside GD3 and nucleotides (Fig. S8). This population of defects are likely the ones that are induced by air plasma treatment of hBN, as evidenced by a 5-fold increase in Trp-induced event density on air plasma treated hBN compared to pristine hBN (Fig. S2g). In addition, a group of amino acids, including Val-, Lys- and Gly-activated defects have a lower density, and a broad wavelength distribution, lack of a dominant population. This could be explained by a decrease in interaction due to the small side chains of Val and Gly. Lys, on the other hand, has a large surface area, but lacks a planar structure and electron donating group, resulting in decreased activation. To summarize, our findings indicate at least two distinct defect populations that can be selectively activated by different amino acids. The activation mechanism likely involves the quantum-confined Stark effect [40,41], where the electric field of the bound molecule perturbs the defect states, triggering emission. However, further studies, including correlative characterization of defect structures and activation emission and first-principle calculations are needed to fully elucidate the chemical specificity of defect activation.

Variation in chemical structures of amino acids also led to different event density. As can be seen in Fig. 3c, the amino acids with large aromatic rings, such as Trp and Tyr, had the highest event densities, followed by amino acids with decreasing surface areas (table 2 in SI). Surprisingly, the event density showed a nice linear correlation with the so-called aromaphilicity index ($R^2 = 0.97$), which measures affinity between amino acids and graphene in aqueous condition[42]. A strong adsorption of molecules containing large planar structures is common for many 2D materials due to the structural similarity to the honeycomb lattice[35,37,42]. In the case of hBN, a highly favorable adsorption of positively charged amino acids, such as Lys and Arg, by electrostatic interaction was predicted by molecular dynamics simulations[34,37]. A positive correlation between the surface area of the amino acids and the event density was observed on hBN ($R^2 = 0.86$) when charged molecules are excluded (Fig. S10b). However, we found large discrepancies between event densities and predicted high adsorption density of the molecules for Arg, Lys and Gly (Fig. S10a). We attributed it to a low defect activation rate by positively charged side chains in Arg and Lys, and the lack of side chain in Gly. In short, our results suggest that the affinity of the amino acids to the hBN surface plays a crucial role in determining the molecule density on hBN surface, however the charge and the chemical structure on the side chains may affect the activation rate of the defect emission.

Further, we assess the reproducibility by analyzing defect emissions across multiple areas on different hBN flakes with five features of the emission histogram, including peak wavelength, mean wavelength, full width half maximal (FWHM), skewness and kurtosis. Paired scattered plots showed that peak wavelength and skewness are the key features of the histogram distinguishing emissions activated by different amino acids (Fig. 3d). Here, each data point represents a measurement on a different surface area of hBN. Measurements of identical amino acids stay fairly close, indicating a relatively good reproducibility among different areas of hBN surface. Using t-stochastic neighbor embedding (t-SNE)[43], we observed effective clustering of amino acids using all 5 features, indicating consistent defect activation patterns (Fig. S11). Such reproducibility is crucial for developing hBN-based bio-sensing applications.

**Switch on defect emission by single deprotonation of amino acids**

The charge state of the amino acid side chains plays a critical role in activating defect emissions, as evidenced by Lys in Fig. 3. By varying the buffer pH, we deliberately switched on defect activation by deprotonating the side chain. We adjusted the buffer pH beyond the pKa values of alpha-amine and side-chain amine in Lys, deprotonating alpha-amine at pH 9.6 and the side chain at pH 11.1. Met, which has uncharged side chain, and Pro, whose side chain shares alpha-amine group, were studied as reference samples under the same conditions. As shown in Fig. 4, the charge state of amino acids significantly influences their interaction with hBN, impacting both event density and emission wavelength. Increasing the pH from 7.4 to 11.1 resulted in a three-fold increase in event density for Lys, accompanied by a new emission population around 600 nm (Fig. 4). In contrast, Pro and Met exhibited a two-fold decrease in event density, with Pro showing an increasing population around 600 nm, whereas Met showed negligible differences. These shifts are attributed to the changes in the charge state of the amino acids, as the defect response to PBS buffer across different pH values remained consistent (Fig. S9)

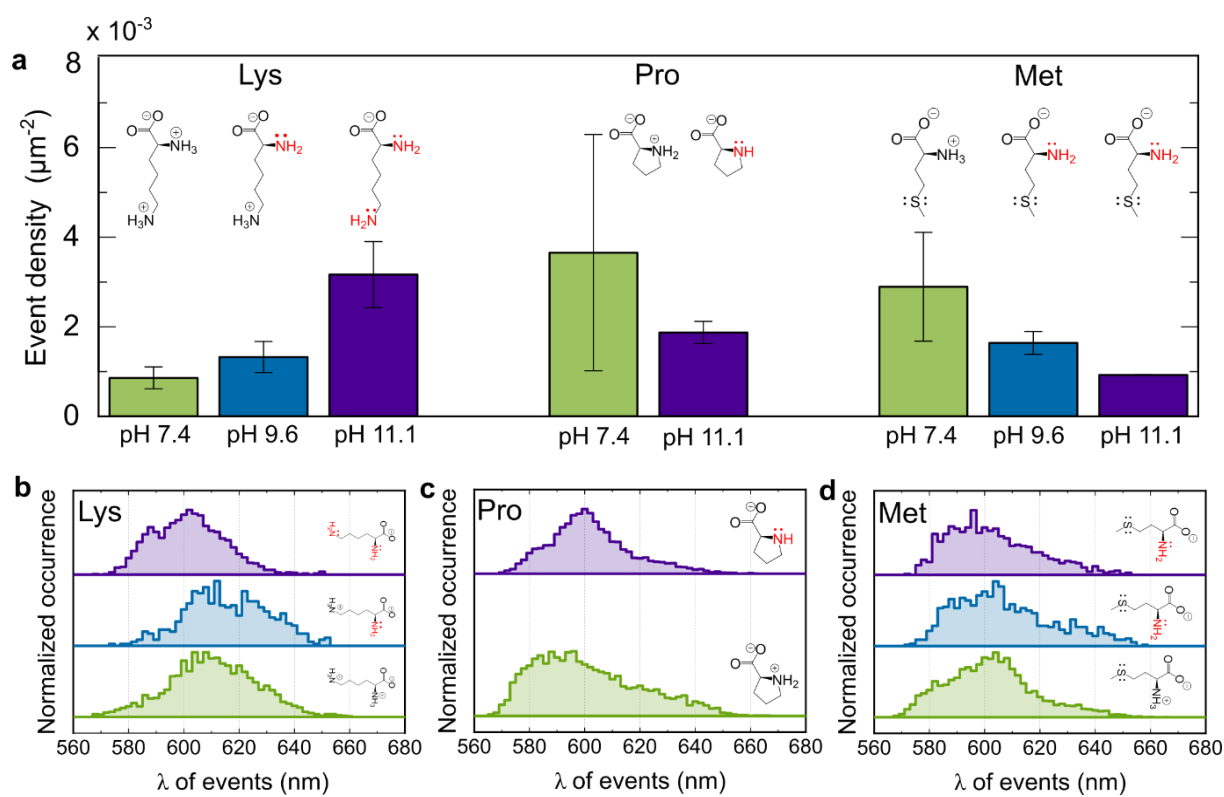

**Figure 4. Charge state of amino acids affects activation of defects. a**, Event rate of lysine, proline, methionine at pH 7.4, 9.6 and 11.1, respectively. **b-d**, Histograms of wavelengths of defects emission activated by lysine, proline, methionine at pH 7.4 (green), pH 9.6 (blue) and pH 11.1 (purple), respectively.

The reduction in event density for Pro and Met is likely due to electrostatic interactions. As pH increases, deprotonation of the alpha-amines causes Pro and Met to become negatively charged, leading to their repulsion from the hBN surface. In contrast, deprotonation of the side chain amine in Lys enabled activation of new defect populations, resulting in an increased event density. The emission profiles of Lys and Pro at pH 11.1 resemble those of Met, indicating that the defect activation around 600 nm is likely driven by electron acceptance from electron-donating groups at their side chains, particularly those with lone pair electrons. Interestingly, previous MD simulations have shown an excess of hydrogen-bond electron acceptors at the water/hBN interface, suggesting that hBN could act as an electron donor[37]. This electron transfer from hBN to water may explain the quenching of defect emissions in aqueous solutions. When molecules with electron donating groups bind to the defects, the defect may accept electrons from these molecules, perturbing electron clouds and activating PL emissions. Further computational calculations using density functional theory may help to unravel the precise mechanism of defect activation. In summary, our findings demonstrate that defect emissions are sensitive to charge state of molecules. By controlling the charge state of biomolecules, it is possible to selectively switch defect emissions in hBN, offering an additional handle for developing a hBN-based biosensor.

## Machine-learning assisted identification of individual amino acids

The ability of hBN to respond with a distinct PL emission to small changes in chemical structure of biomolecules opens the way for the sensing and single-molecule detection of biomolecules. We investigated the ability to identify individual amino acids based on single-event information, employing a Random Forest classification[44] model as a proof of concept (Fig. 5a). A Random Forest classification model is an ensemble learning method that combines multiple decision trees to improve prediction accuracy and reduce overfitting [44].

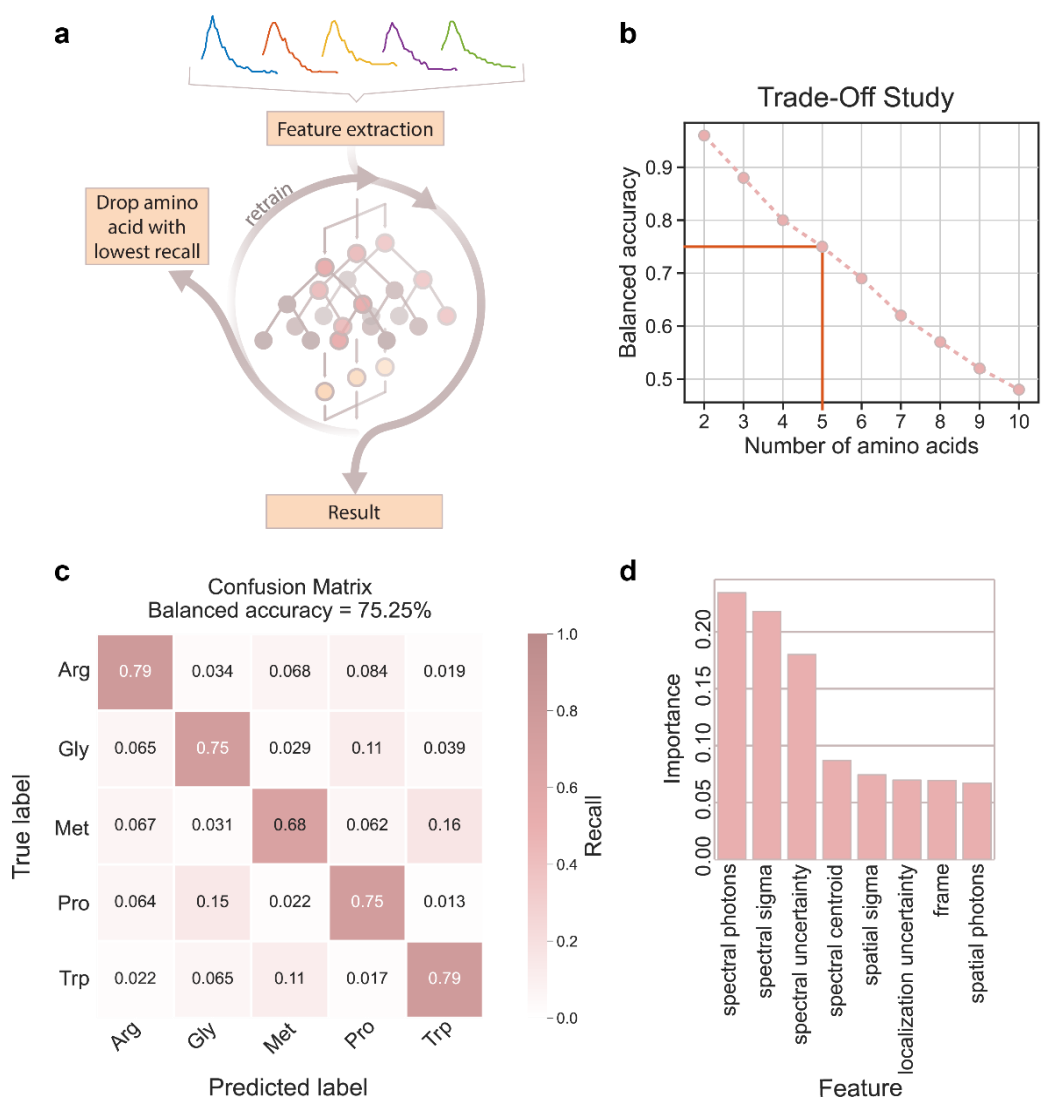

**Figure 5. Analysis of Random Forest classification on amino acids. a**, Model training workflow. **b**, Trade-off analysis between the model accuracy and the number of amino acids identified. Tested on models trained with decreasing number of amino acid classes by removing the lowest recall amino acid class each iteration. **c**, Confusion matrix of the selected model, classifying 5 amino acids. The model prediction (x-axis) is compared to the ground truth (y-axis) and the number of correct predictions is represented by the recall score. **d**, Feature importance plot for the five-amino-acid model, showing the contribution of each feature to predictions, with importance values summing to 1.

For our analysis, we first trained a Random Forest classification model on single-event data for all ten amino acid classes. Taking advantage of the information-rich data acquired in sPAINT, features of single molecule events, including spatial photons, localization uncertainty, spatial sigma, spectral photons, spectral centroids, spectral sigma, spectral uncertainty, and frames were extracted (Methods). The features dataset was then randomly split into a training set (80%), a validation set (10%), and a test set (10%). To address class imbalance[45], we under-sampled the over-weighted classes so all amino acid classes in the training set would have the same number of observations and saved the removed observations in an additional test set to avoid wasting data. From the resulting model, we assessed the recall (true positive rate) for each amino acid class and removed the class with the lowest recall (Table 3 in SI), to evaluate the trade-off between the number of amino acids the model could identify simultaneously and the classification accuracy (Fig. 5b). The balanced accuracy was close to 50% for identification of ten amino acids and ramped up to 75% for a selection of five amino acids.

This trade-off study allows for the selection of a model that best suits specific needs: a model with higher accuracy but fewer amino acids identified, or one with lower accuracy but a greater number of amino acids identified. For this proof of concept, we have selected the Random Forest model that classifies five amino acids with an accuracy of 75%, which we deemed suitable for general applications. This accuracy is significantly higher than the 20% expected from random chance, demonstrating great potential of this technique for label-free single-molecule identification. The confusion matrix in Fig. 5c shows the five most distinguishable amino acids chosen by model, including Trp, Arg, Med, Pro and Gly, with their recall scores. The recall scores, indicating the proportion of correct predictions relative to the total predictions made for the specific amino acids, range from 0,68 to 0,79. The five chosen amino acids cover a large variety of chemical structures in terms of the side chains, which are in line with the spectral observations as discussed in the previous section (Fig. 3).

Lastly, we analyzed the feature importance for the five-amino-acid model (Fig. 5d), which highlights the contribution of each feature to the predictions represented in the confusion matrix. The three most important features include spectral photons, spectral sigma and spectral uncertainty, indicating that the intensity of events and the spectral shape play a crucial role for identifying amino acids. These features are also correlated with the emission wavelength. Manual identification based on single features or 2D scatter plots of paired features proved to be much less effective, only being able to separate the two most different amino acids effectively (Fig. S14). This limitation underscores the strength of combining multidimensional microscopy and machine learning analysis, which excels at detecting complex patterns in high-dimensional data and can integrate multiple correlated features to improve classification accuracy.

**Conclusions**

Our exploration of hBN's surface defects for selective biomolecule detection opens new pathways for understanding molecular interactions at the single-molecule level in aqueous environments. By harnessing the wealth of information from sPAINT single molecule microscopy, the unique spectral fingerprints arising from the interaction between hBN defects and various biomolecules can be extracted. We have demonstrated the ability to differentiate between various types of lipids and amino acids with remarkable precision. This selective activation of defects, observed through spectrally-resolved super-resolution microscopy has allowed us to detect subtle chemical differences, such as single deprotonations, and to capture the dynamic conformational changes of biomolecules on the hBN surface.

Moreover, by combining multidimensional microscopy and machine learning, five distinct amino acids were identified based on their spectral signatures with over 75% accuracy, which underscores the potential of this technique for label-free biomolecule identification. These findings not only advance our understanding of hBN-biomolecule interactions but also establish a promising platform for future research in molecular diagnostics and biosensing. Moving forward, we foresee our approach paving the way for the development of sophisticated, label-free detection systems that could revolutionize the study of complex biological systems and facilitate new diagnostic technologies.

## Methods

### Chemicals used and their abbreviations

DPPA: 1,2-dihexadecanoyl-sn-glycero-3-phosphate, DPPE:1,2-dihexadecanoyl-sn-glycero-3-phosphoethanolamine, DPPC: 1,2-dihexadecanoyl-sn-glycero-3-phosphocholine, DOPC: 1,2-di-(9Z-octadecenoyl)-sn-glycero-3-phosphocholine, 14:1 PC: 1,2-di-(9Z-tetradecenoyl)-sn-glycero-3-phosphocholine, GD3: ganglioside GD3, TRP: tryptophan, TYR: tyrosine, ARG: arginine, His: histidine, PHE: phenylalanine, MET: methionine, PRO: proline, VAL: valine, LYS: lysine, GLY: glycine, dAMP: deoxyadenosine monophosphate, dGMP: deoxyguanosine monophosphate, dCMP: deoxycytidine monophosphate, dTMP: deoxythymidine monophosphate.

### Preparation of hBN samples

hBN flakes were exfoliated from commercially available crystals (HQ Graphene) by low adhesion tape (BT-130E-SL, Nitto) and transferred to round glass coverslips with 50 µm grids (#1.5H D263 Schott glass, ibidi). First, the coverslips were immersed in 2% Hellmanex III for two hours and rinsed by ultra-pure water to clean the surface. The coverslips were then treated in atmospheric pressure air plasma operated at 280 V, 15.6 A, 21kHz (FG5001+RD1004, Plasmatreat) for 1 min to increase the adhesion of the surface before hBN transfer. To induce defects, hBN transferred on coverslips were treated in the same air plasma for 10 seconds. The coverslips were then mounted on a fluidic chamber (Attofluor cell chamber, Invitrogen) for imaging on the microscope.

### Preparation of biomolecule samples

Lipid dry powder (DPPA, DPPE, DPPC, DOPC, 14:1 PC, ganglioside GD3 from Avanti Polar Lipids) was firstly dissolved in chloroform at a concentration of 5 mM in a glass vial. The stock solution was then evaporated with $N_2$ flow on a vortex mixer to form a lipid film on the interior of the glass vial. The rehydrate the lipids with PBS buffer. For lipids which have a melting temperature higher than the room temperature, PBS buffer was heated above the melting temperature before and during the rehydration process. Vortex the solution to redissolve all lipid materials and to form multilamellar vesicles. The stock solution were stored in 4°C. Before experiments, dilutions were made with concentrations ranging from 20 µM – 3mM from the stock solution.

All amino acid and nucleotide samples were freshly prepared before experiments. Dry powder of amino acids and nucleotides were weighted and dissolved in PBS buffer to obtain the desired concentrations. Typically measurements of amino acids were done at 1 mM and 5 mM; nucleotides at 5 mM.

### sPAINT microscopy

hBN defects emissions were imaged on an inverted TIRF microscope (Nikon Ti) through an oil-immersion objective lens (Nikon Apo TIRF 100x Oil DIC N2) with a numerical aperture of 1.49. Excitation laser with wavelength of 532 nm (FP1280764, coherent OBIS) was focused on the back focal plane of the objective lens to achieve a wide-field illumination. The laser power is set at 50 mW unless otherwise stated. The emissions were collected by the same objective lens and expanded by a 1.5x relay lens before passing through a slit (TwinCam, Cairn) for cutting the field of view. The emission signal went through a notch filter (NF533-17, Thorlabs) and a long-pass filter (ET542LP, Chroma Technology USA) that cuts off at 537 nm to allow us probing emissions of wavelengths from 540 nm to 750 nm. An transmission grating (70 grooves/mm, 25x25 mm-46-068, Edmund) was installed ~2cm before the EMCCD camera (Andor DU-888 X-9414) to split 41% signal to the $0^{th}$-order diffraction and 32% into $1^{st}$ order diffraction image (Fig. 1a). Using a mercury wavelength calibration lamp (HG-2, Ocean Optics), spectral wavelengths and pixel positions were correlated. During imaging, camera was set as electron multiplication gain of 250, exposure time of 50 ms, unless otherwise specified. Typically 5.5k frames were recorded for lipid and nucleotide samples, 22k frames for amino acids.

**Image processing**

The image stacks were processed using RainbowSTORM[46] and ThunderSTORM[47] plug-ins in Fiji. In the spatial image (0$^{th}$-order diffraction), the coordinates of emission events (x, y, t) were extracted by fitting the point spread function. The localization of emission events were achieved with an uncertainty of $\delta = \sigma/\sqrt{N}$, where $\sigma$ is the width of the point spread function defined by the optical system, $N$ is the number of photons in the 0$^{th}$-order diffraction image. The localization uncertainty ranged from 15 to 25 nm in our measurements. The coordinates of a fixed emitter follow a 2-D Gaussian distribution covering an area with diameter of 6×δ. The coordinates were then used to calculate event density per frame and connected between consecutive frames within a set searching diameter for kinetic and dynamic analysis by using a MATLAB script (details in SI). To extract the association time of a molecule binding on a defect, a searching diameter of 6×δ was set. Whereas for tracking the lipid diffusion, searching diameters in the range of 200 nm to 600 nm were used depending on the diffusivity of the molecule. In the spectral image (1$^{st}$-order diffraction), the wavelength of the dispersed light from each emission event were calculated based on the calibration curve that relates the wavelength in nanometer to the spatial distance in pixels. Features of spectrum, including spectral photons, spectral centroid (λ of event, mean wavelength of the spectrum), spectral sigma (spectral width extracted by Gaussian fitting), spectral uncertainty were extracted for each events. Detailed descriptions of the features can be found in previous report[46].

**Machine Learning Classification**

The Random Forest Classification models were trained using Scikit-Learn's RandomForestClassifier[48]. The chosen five-amino-acid model was trained with the following hyperparameters: [max_depth = 30, max_features = None, min_samples_leaf = 2, n_estimators = 1400]. All models in the trade-off study were optimized using the Scikit-Learn's class RandomizedSearchCV[49] for 100 different parameter settings on 3-fold cross validation, using all processors available. After the optimization, the best estimator was selected to proceed with the analysis and obtain the confusion matrix and classification metrics. The models were trained and optimized on an AMD Ryzen 9 5900X 12-Core Processor, 3701 Mhz, 12 Cores, and 24 Logical Processors, running on Windows 10 Enterprise.

**ACKNOWLEDGEMENTS**

We acknowledge Nathan Ronceray and Ilya Sychugov for insightful discussions. We are grateful to Chan Cao and Louis Perrin for their effort in setting up MD simulations. We also thank Emanuele Criscuolo for making the schematics of amino acids and hBN in PyMol. M.Z. acknowledges the financial support from the Marie Sklodowska-Curie Individual Fellowship (101068746-2DMAP).

**AUTHOR CONTRIBUTIONS**

M.Z. and L.A. conceived the projects. M.Z. designed and performed the experiments. S.V. prepared the lipid samples. M.Z. analyzed and interpretated the data. C.I.L. performed machine learning. M.Z. wrote the manuscript with input from all authors.

**COMPETING INTERESTS**

The authors declare no competing financial interests.

# Supplementary Information for Label free identification of biomolecules by single-defect-spectroscopy at the aqueous hexagonal boron nitride interface

Miao Zhang*, Cristina Izquierdo Lozano, Stijn van Veen , and Lorenzo Albertazzi*

Department of Biomedical Engineering, and Institute for Complex Molecular Systems, Eindhoven University of Technology, 5600MBEindhoven, The Netherlands

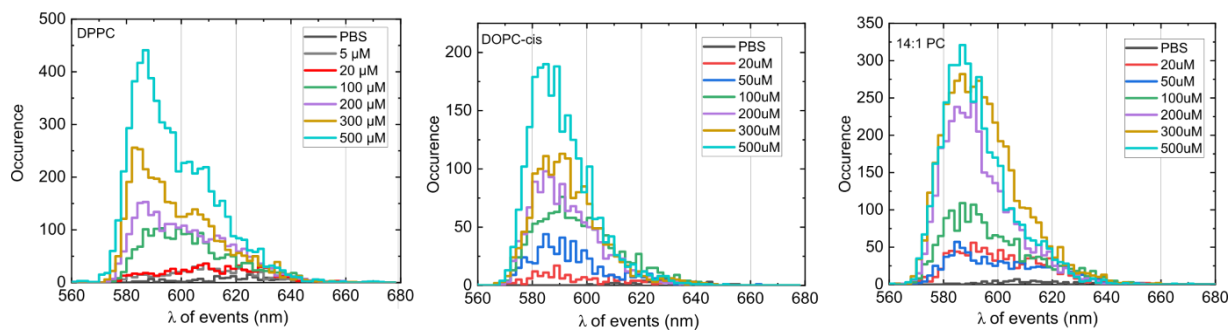

Figure S1. Histogram of emission wavelengths activated by DPPC, DOPC, and 14:1 PC lipid molecules at various concentrations on identical surface area of hBN. The numbers of emissions activated by the molecules are all above the level observed in PBS.

**Defects in hBN**

Defect emissions were activated by pure acetonitrile (ACN). Two populations of bright emissions were observed around 580 nm and 640 nm, respectively (Fig. S2a-c). However, additional water in ACN quenched the most of emissions, especially the population around 640 nm. Air plasma altered defect structure and induced additional defects in hBN surface. This was evidenced in acetonitrile activated defects emissions on hBN treated with air plasma of different time periods (Fig. S2e). The defect population around 640 nm were destroyed after just 10 sec of air plasma treatment. Whereas defect populations between 600-640 nm were increased by air plasma treatment. Air plasma treatment longer than 10 seconds did not induce significant changes in defect populations. DPPC activated defect emission on pristine and air plasma treated hBN dominated between 580-600 nm, showing negligible changes(Fig. S2f). Whereas, tryptophan activated defect emission around 630 increased 5-fold when compared between air plasma treated hBN and pristine hBN (Fig. S2g).

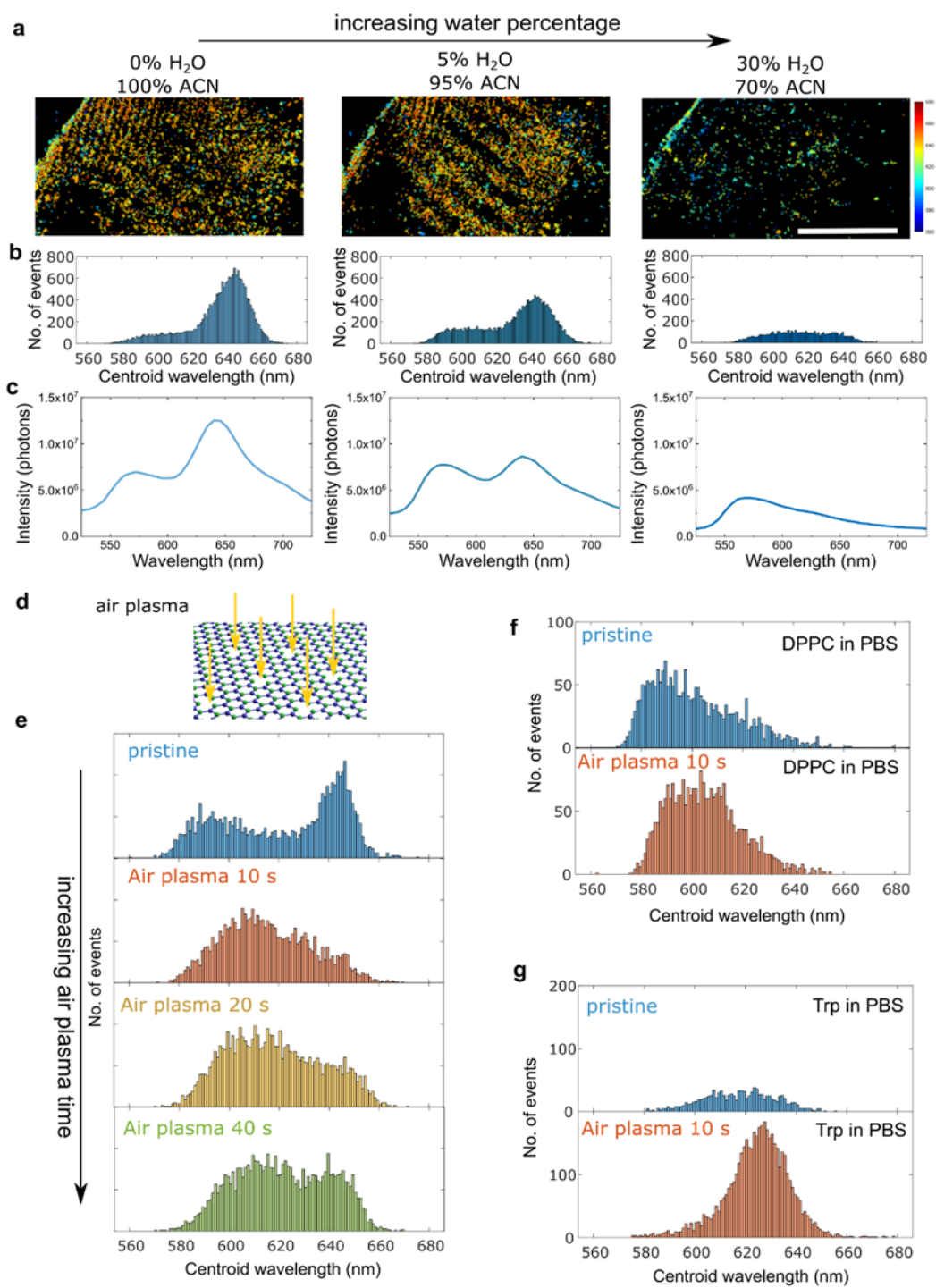

Figure S2. Intrinsic defects in pristine hBN versus air plasma induced defects. **a**, Maps of defect emissions activated by acetonitrile (ACN)/water mixture with an increasing ratio on pristine hBN surface. **b** and **c**, Histograms of event wavelengths and ensemble spectra of defects emissions according to the conditions in **a**. **d**, an sketch of defects created by air plasma. **e**, Histograms of event wavelengths activated by acetonitrile on hBN surfaces treated with increasing time of air plasma. **f**, Histograms of event wavelengths activated by DPPC in PBS (100 μM) on pristine and air plasma treated hBN. **g**, Histograms of event wavelengths activated by tryptophan(Trp) in PBS (1mM) on pristine and air plasma treated hBN.

**Chemical specificity of molecule activated defect emissions**

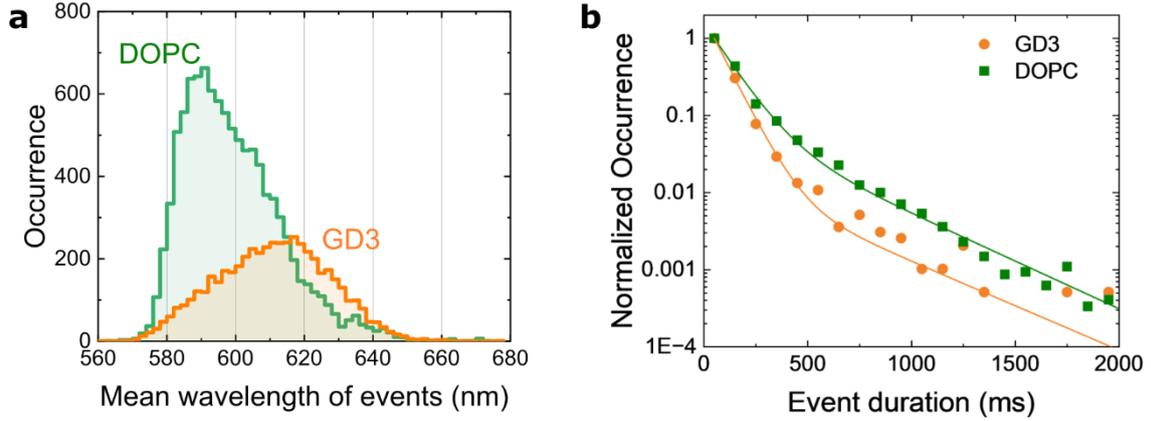

Figure S3. Emissions caused by different molecules on air plasma treated hBN. **a**. Histogram of mean wavelength of events activated by DOPC and ganglioside GD3 lipids. **b**. Histogram of durations of emissions activated by DOPC and ganglioside GD3. The data was fitted by double exponential decay function.

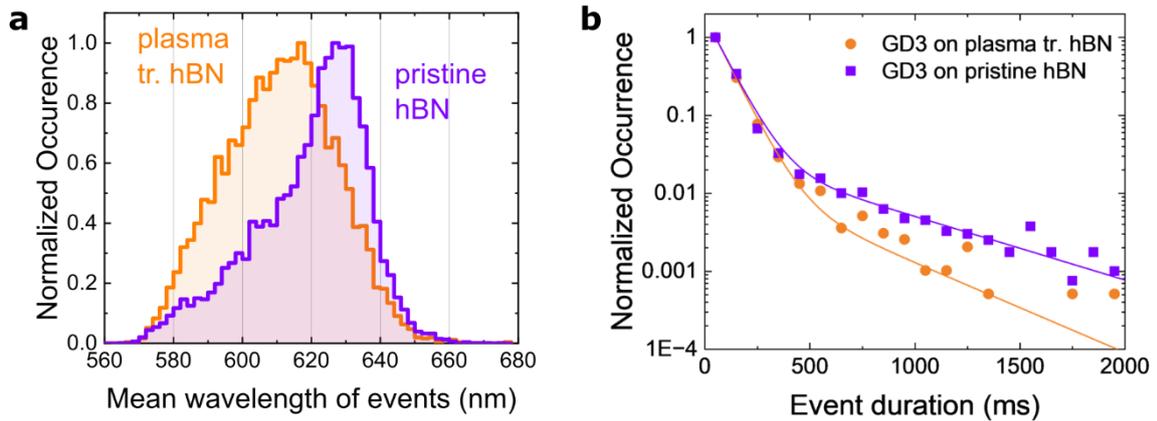

Figure S4. Comparison of events caused by ganglioside GD3 on air plasma treated hBN and on pristine hBN. **a**. Histogram of mean wavelength of emissions activated by ganglioside GD3 lipids on air plasma treated hBN and on pristine hBN. **b**. Histogram of durations of emissions activated by ganglioside GD3 on plasma treated hBN and pristine hBN. The data was fitted by double exponential decay function.

**Adsorption isotherms of DOPC and 14:1PC**

The adsorption isotherms of DOPC and 14:1 PC are described by a general surfactant adsorption isotherm equation (eq 1), which accounts for a 2 step adsorption model: i) monomer adsorption without interaction with each other; ii) fast adsorption induced by hydrophobic interaction between monomers.[1,2]

$$\Gamma = \frac{\Gamma_\infty k_1 C(\frac{1}{n} + k_2 C^{n-1})}{1 + k_1 C(1 + k_2 C^{n-1})} \qquad (1)$$

Where $\Gamma$ is the adsorption density, $\Gamma_\infty$ is the adsorption density at saturation, $C$ is the concentration of surfactant, $k_1$ is the equilibrium constant of the first step, $k_2$ is the equilibrium constant of the second step, $n$ is the aggregation number.

Fitting the adsorption curve with the equation is not feasible due to the high degree of freedom. We generate the adsorption curve by firstly extracting $k_1$ by fitting the low concentration regime and extracting $\Gamma_\infty$ from the saturation regime. Then $k_2$ and $n$ were carefully selected to generate adsorption isotherm that matches with the experimental data. The parameters for DOPC and 14:1 PC are listed in the Table.1 below.

*Table 1*

|        | $k_1(\mu M^{-1})$ | $k_2(\mu M^{-1})$ | $n$ | $\Gamma_\infty(\mu M^{-1}s^{-1})$ |
|--------|-------------------|-------------------|-----|----------------------------------|
| DOPC   | 0.0017            | 0.006             | 2   | 0.28                             |
| 14:1 PC| 0.002             | 0.04              | 2   | 0.1                              |

**Analysis of dynamics of lipid on hBN surface**

Dynamics of lipid binding/unbinding to the defects and hopping between defects was analyzed by connecting localizations of emission events in consecutive frames within a set distance limit ($d_{max}$). To extract the association time (event duration of a lipid binding to a single defect), the distance limit was set as 6 × *localization uncertainty*, to include 99.7% localization events from a single emitter. For recordings of 20 frame per second (fps), localization uncertainty for DOPC and DPPC shown in Fig. SX. We set dmax = 150 nm , 150 nm and 120 nm for 14:1 PC, DOPC and DPPC, respectively, to extract association time of a molecule to a single defect.

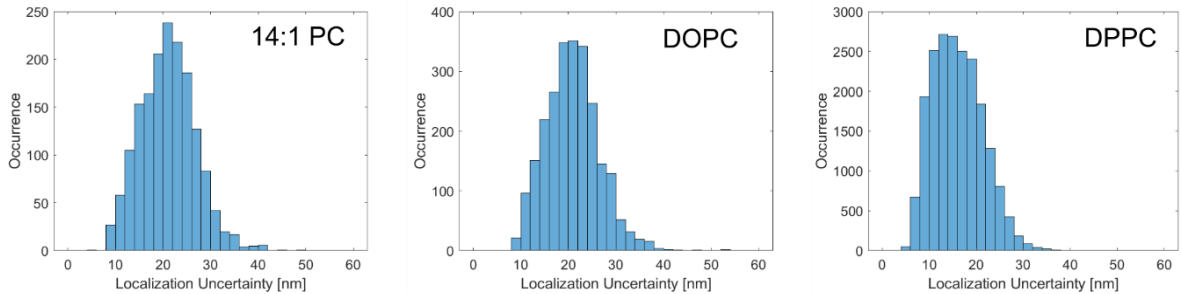

Figure S5. Localization uncertainty of 14:1 PC, DOPC and DPPC activated defects emission.

To extract the trajectories of lipid hopping between defects, localizations in consecutive frames within a distance limit ($d_{max}$) were connected. We estimated the lower limit of $d_{max}$ by diffusion coefficient in literature. Previously measured lateral diffusion coefficients of lipids in supported lipid bilayers by fluorescence recovery after photobleaching (FRAP) showed that DOPC is in the range of 1-10 μm$^2$ s$^{-1}$, whereas DPPC is in the range of 0.05-0.1 μm$^2$ s$^{-1}$.[3,4] For recording at 20 frame per second, the lower limit of $d_{max}$ is 700 nm for DOPC and 70 nm for DPPC. For recording at 48 frame per second, the lower limit of $d_{max}$ is 460 nm for DOPC and 46 nm for DPPC. The upper limit of $d_{max}$ is defined by the event density. After carefully checking the individual mean square displacement curves, for recordings of 20 fps, we set the $d_{max}$ = 600 nm for DOPC and 14:PC, $d_{max}$ = 300 nm for DPPC; whereas for recordings of 48 fps, we set $d_{max}$ = 600 nm for DOPC and 14:PC, $d_{max}$ = 200 nm for DPPC. The analysis was done by a custom-made MATLAB code, which utilizes a part of the particle tracking code by Daniel Blair and Eric Dufresne (https://site.physics.georgetown.edu/matlab/code.html) and a part of the diffusion analysis code by Jean-Yves Tinevez (http://tinevez.github.io/msdanalyzer/).[5] Localization data extracted by Fiji Plugin Thunderstorm was used as input data.

To record the fast dynamics of lipid hopping between defects, we removed the grating to boost the number of photons in images, and cropped the field of view to achieve a faster read out of the EMCCD camera. Image stacks of lipid diffusion on hBN surface at 1 mM were recorded at 48 fps. The trajectories of DOPC, 14:PC and DPPA diffusion are shown in Fig. S6. DOPC and 14:1 PC clearly had longer trajectories compared to DPPA. DPPA were confined at single spots. Quantitative track analysis confirmed the observation. While DOPC and 14:1 PC have shorter track times than DPPA, their displacement, total distance travelled and the max distance travelled are significant longer than DPPA. The diffusion coefficient extracted from 48 fps recording are 0.71 $\mu m^2 s^{-1}$ for DOPC, 0,54 $\mu m^2 s^{-1}$ for 14:1 PC and 0.025 $\mu m^2 s^{-1}$ for DPPA. This agrees well with the diffusion coefficients that were measured by FRAP.[3,4]

**Discussion on how phase of lipid membrane affect the event density**

Lipid membranes at different phases have different packing orders. DPPC at solid-gel phase pack highly ordered with an area per lipid of 47.9 Å$^2$,[6] whereas DOPC at liquid-disordered phase packs with disordered hydrocarbon chains with an area per lipid of 72.2 Å$^2$.[7] An increased area per lipid reflects that the hydrocarbon chains have a higher degree of freedom in conformational change. This also leads to an increased water permeability[8], which negatively influence the defect activation. This agrees well with our observations of lower density of defects activated by DOPC compared to DPPC.

Adding cholesterol (1:1 ratio) to DPPC results in phase transition from the solid-gel phase to the liquid-ordered phase, increasing the area per lipid to 55 Å$^2$ at room temperature.[9] This expanded area per lipid is associated with lower chain order and higher water permeability[8]. These could lead to a lower probability of the binding of hydrocarbon chain to the defect and enhanced quenching by water at the interface. Together, these factors lead to a lower defect activation rate. In contrast, adding cholesterol to DOPC at liquid-disorder phase slightly increases the packing order with a decreased area per lipid of 64 Å$^2$.[10] However, despite the change in packing, we did not detect a corresponding increase in defect activation density in the DOPC/Cholesterol sample. This lack of increase may be due to the fast diffusion of DOPC , which limits the duration of bound events, preventing sufficient interaction with defects regardless of cholesterol presence.

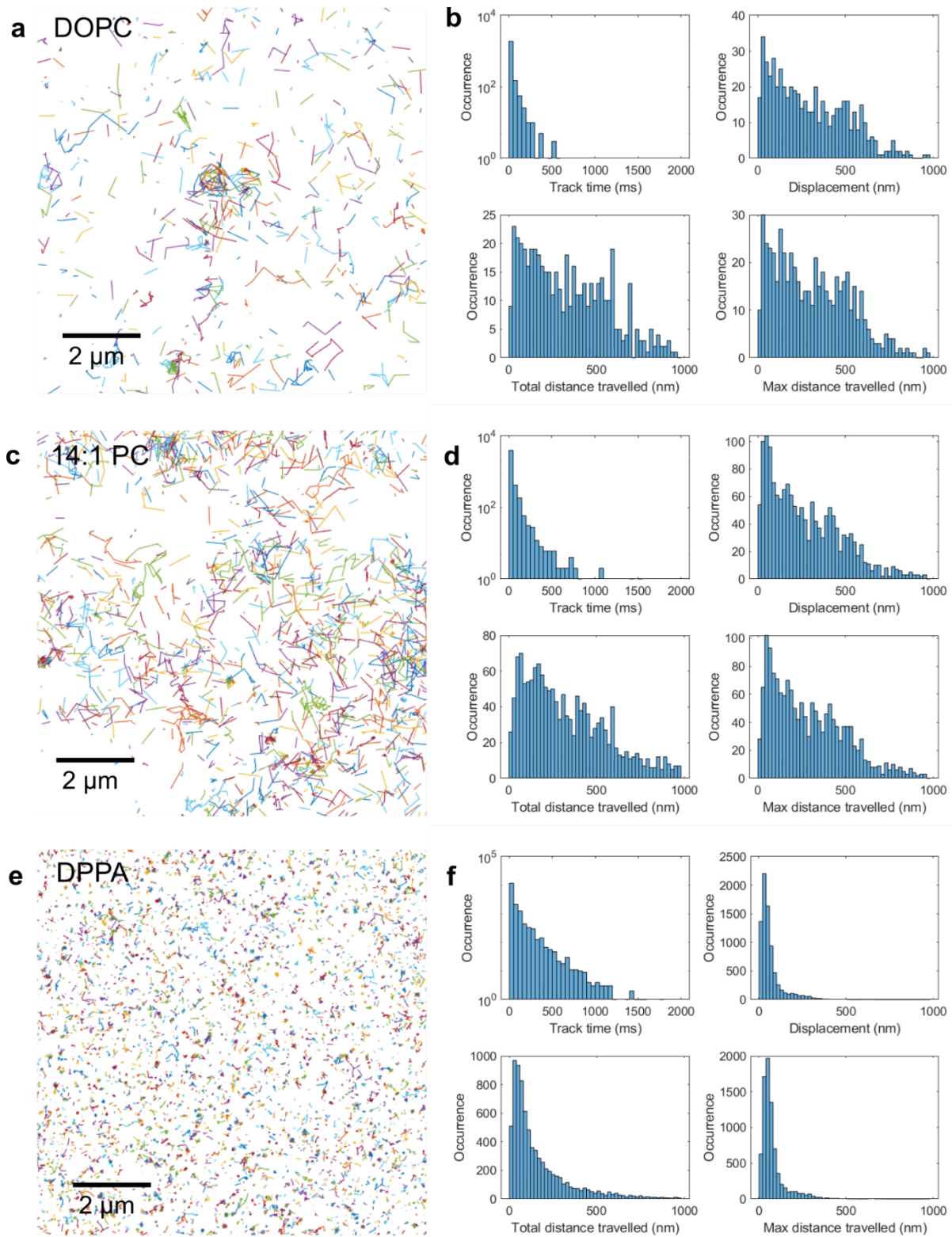

Figure S6. Diffusion behavior of DOPC, 14:PC and DPPA at 1 mM. **a,c,e** trajectories of lipid diffusion on hBN surface. **b,d,f** trajectory information, including histogram of track duration, displacement, total distance travelled, and max distance travelled per trajectory.

**Staining lipids by CellMask orange**

Bright emissions on hBN were observed in the presence of 0.55mM DOPC, as can be seen in Fig. S7a. The reconstructed super-resolution image shows emission events fully cover the hBN surface. To confirm lipid presence on the hBN surface, we used the lipophilic dye (CellMask orange, Invitrogen, 0.5 μg/mL) for staining. As can be seen in Fig. S7c, significant staining occurred on hBN surface. Interestingly, the hBN surface in pure PBS was also stained (Fig. S7d), suggesting a hydrophobic nature of the hBN surface. Nevertheless, the staining intensity on hBN in the presence of DOPC was four-fold higher than that on the bare hBN surface (Fig. S7e), indicating accumulation of lipids on the hBN surface. Note that the thick stripes on hBN surface are caused by interference of the 561 nm excitation laser in relatively thick hBN.

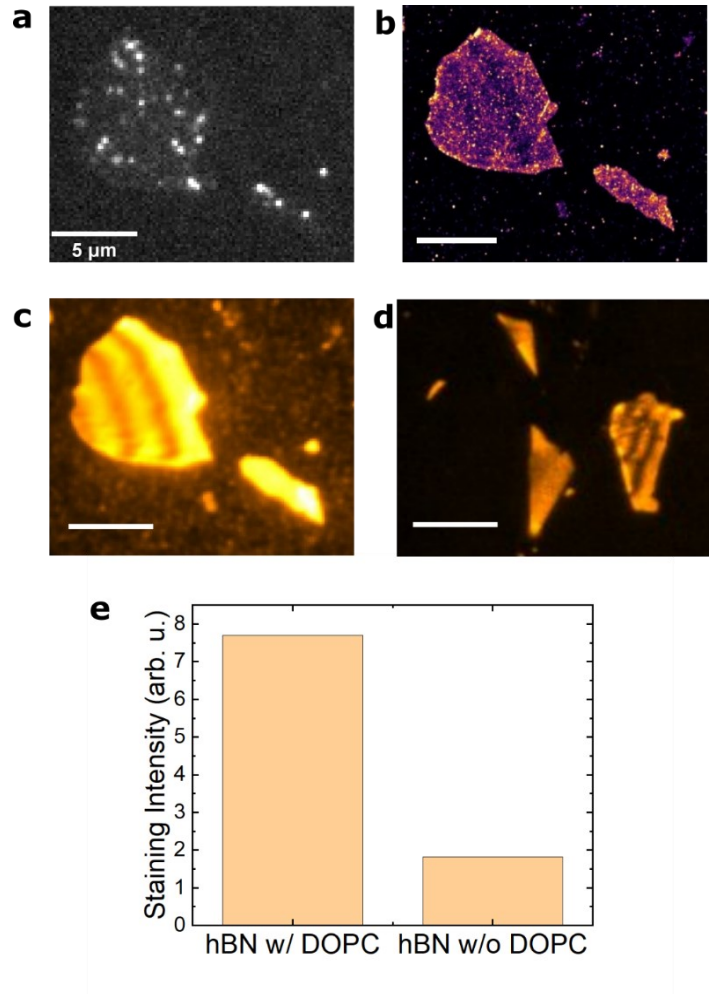

Figure S7. **a**, photoluminescence image of DOPC-activated defects emission on hBN crystals. b, Reconstructed super-resolution image of defect emissions. **c**, a fluorescence imaging of the same area stained by lipophilic dyes (cellmask orange). **d**, a fluorescence image of bare hBN stained by cellmask orange. Scale bar: 5μm. Excited by 561 nm laser. **e**, Averaged fluorescence intensity of cellmask orange staining on hBN with or without DOPC.

**Defect emission activated by nucleic acids**

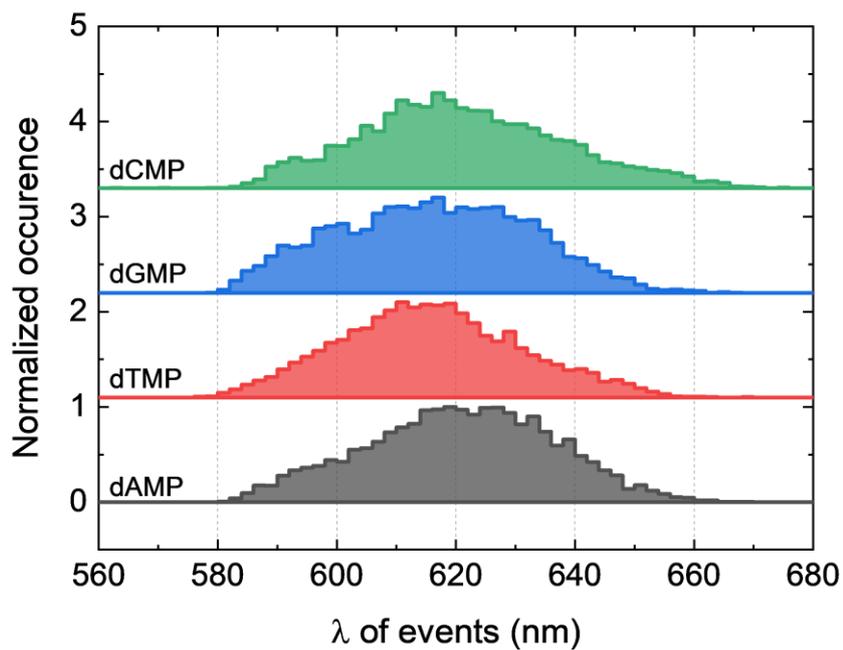

Figure. S8 Normalized histograms of defect emissions activated by nucleic acids, including dAMP, dTMP, dGMP, and dCMP in PBS buffer on air plasma treated hBN surface.

**Defect emission in PBS at different pH values**

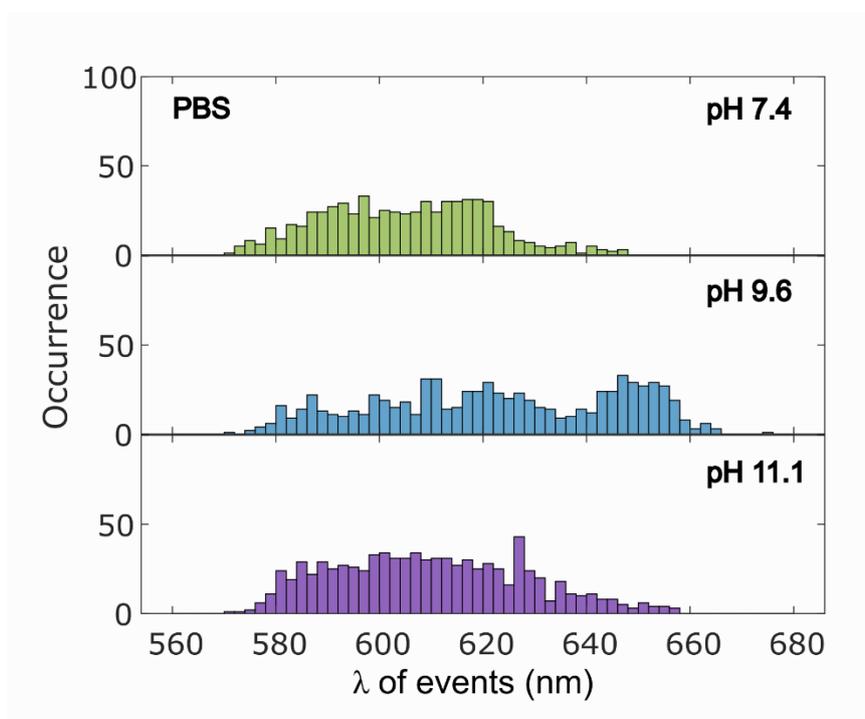

Figure. S9 Histograms of defect emissions activated by PBS at different pH values.

**amino acids activated defect emissions**

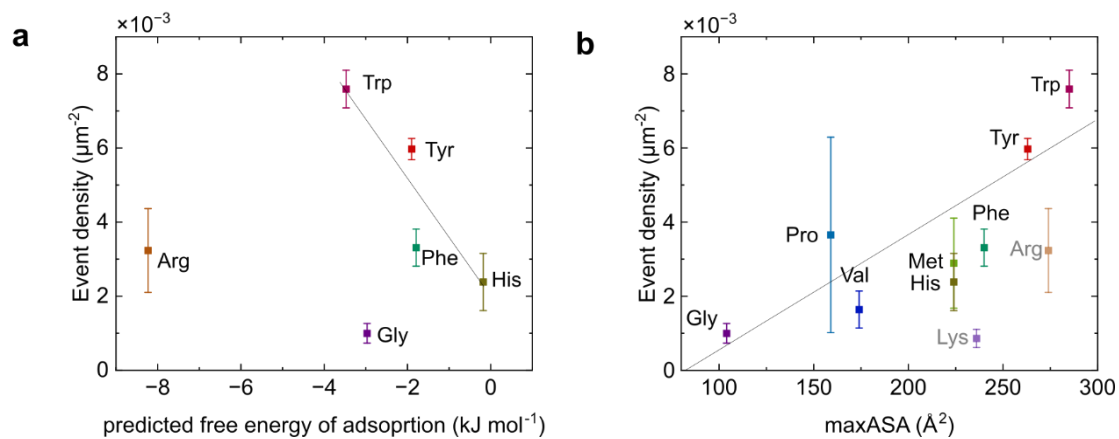

Figure S10. **a**, Event density of defect emission as a function of the predicted free binding energy for amino acids at the aqueous hBN interface. The free binding energy was extracted from the literature.[11] Our results showed that the event density had a linear correlation to the predicted free binding energy for amino acids with uncharged planar side chains, Trp, Try, Phe, His, on hBN surface, while Arg with positively charged side chain and Gly with no side chain, fall as outliers. **b**, Event density as a function of maximal allowed solvent accessibility.[12] A high correlation between event density and maxASA was obtained ($R^2 = 0.86$) when Lys and Arg were excluded.

*Table 2. Property chart of amino acids*

|  | MaxASA (theoritical)[12] | MaxASA (empirical)[12] | Free binding energy to hBN[11] | Aromaphicility index (affinity to graphene)[13] | $\sigma_I$[14] |
|---|---|---|---|---|---|
|  | Å² | Å² | KJ/mol |  |  |
| Trp | 285 | 264 | -3.473 | 1 | 0.06 |
| Tyr | 263 | 255 | -1.9 | 0.85 | 0.05 |
| Arg | 274 | 265 | -8.23 | 0.75 | -0.26 |
| His | 224 | 216 | -0.18 | 0.45 | -0.01 |
| Met | 224 | 203 |  | 0.325 | 0.08 |
| Phe | 240 | 228 | -1.79 | 0.575 | 0.04 |
| Pro | 159 | 154 |  | 0.125 | 0 |
| Val | 174 | 165 |  | 0.15 | 0.01 |
| Lys | 236 | 230 |  | 0.1 | -0.16 |
| Gly | 104 | 97 | -2.97 | 0 | 0 |

$\sigma_I$ electronic property of side chains of amino acids. $\sigma_I > 0$ means electron donating; $\sigma_I < 0$ means electron accepting.[14]

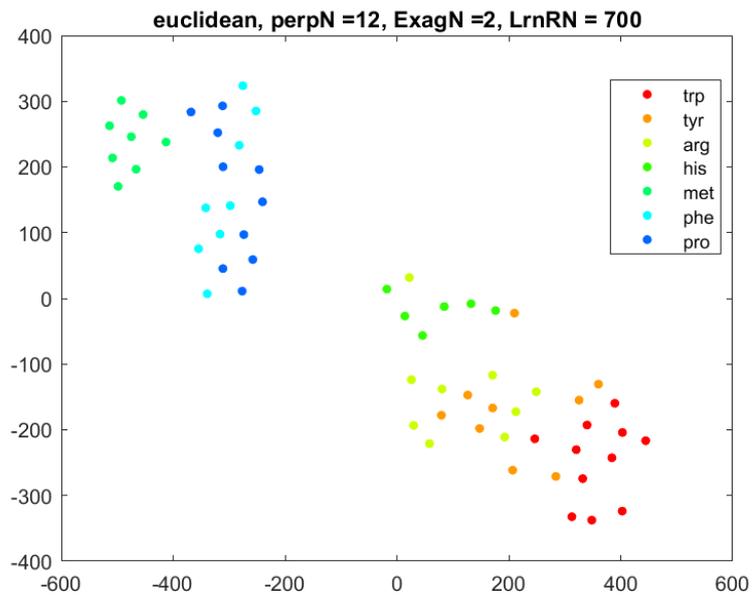

Figure S11. Clustering of amino acids measured on different areas of hBN surface by t-stochastic neighbor embedding (t-SNE) using 5 features extracted from wavelength histograms, including peak wavelength, mean wavelength, full width half maximal (FWHM), skewness and kurtosis. The x and y-axis show unitless numbers.

## Random Forest Classification on features

*Table 3. Order in which the amino acids were removed during the trade-off study, based on their recall score. This order can be related to the "most confused" amino acids when the models are based on their spectral data features.*

|   | **Amino acid removal order:** |
|---|---|
| 1 | Histidine |
| 2 | Phenylalanine |
| 3 | Tyrosine |
| 4 | Valine |
| 5 | Lysine |
| 6 | Methionine |
| 7 | Glycine |
| 8 | Arginine |

## Random Forest Classification on spectral data

The amino acid classification was performed directly on the single defect spectra without extracting the features, as a proof of concept. This could potentially avoid calculation errors and reduce analysis biases, as well as reducing analysis time and effort. As shown in Fig. S12, S13, despite the accuracy being slightly lower, the classification is still successful.

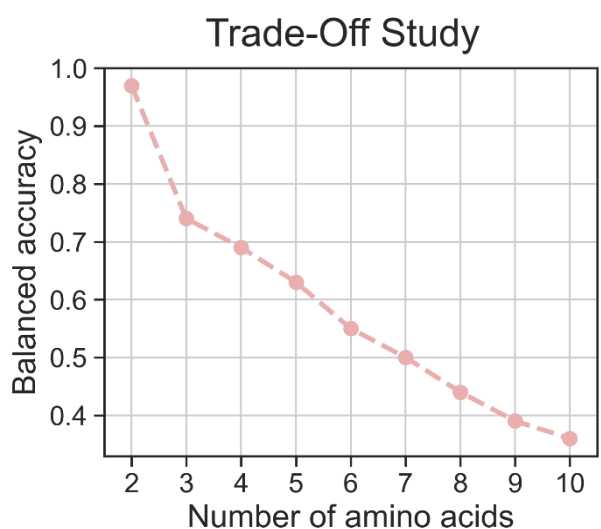

Figure S12. Trade-off study for Random Forest Classification models trained on spectral data vectors.

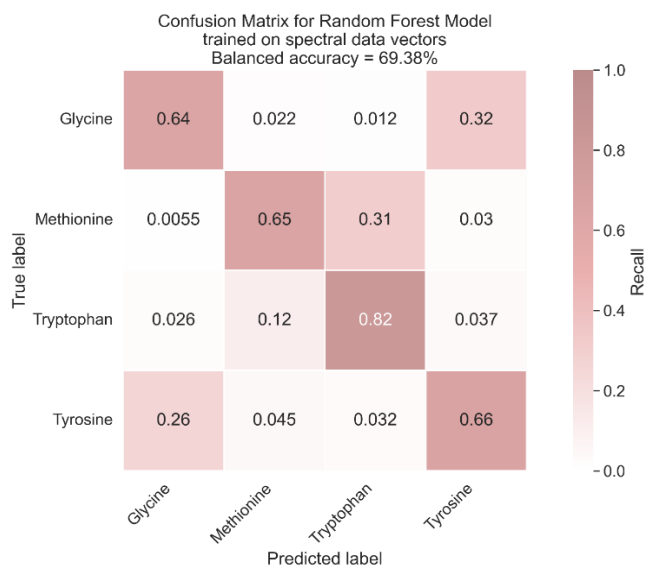

Figure S13. Confusion Matrix for the Random Forest Classification model trained on spectral data vectors. The four amino acids were selected based on their recall during the accuracy vs. classes trade-off study.

Table 4. Order in which the amino acids were removed during the trade-off study, based on their recall score. This order can be related to the "most confused" amino acids when the models are based on the raw spectral data.

|   | Amino acid removal order: |
|---|---|
| 1 | Valine |
| 2 | Histidine |
| 3 | Phenylalanine |
| 4 | Proline |
| 5 | Lysine |
| 6 | Arginine |
| 7 | Methionine |
| 8 | Glycine |

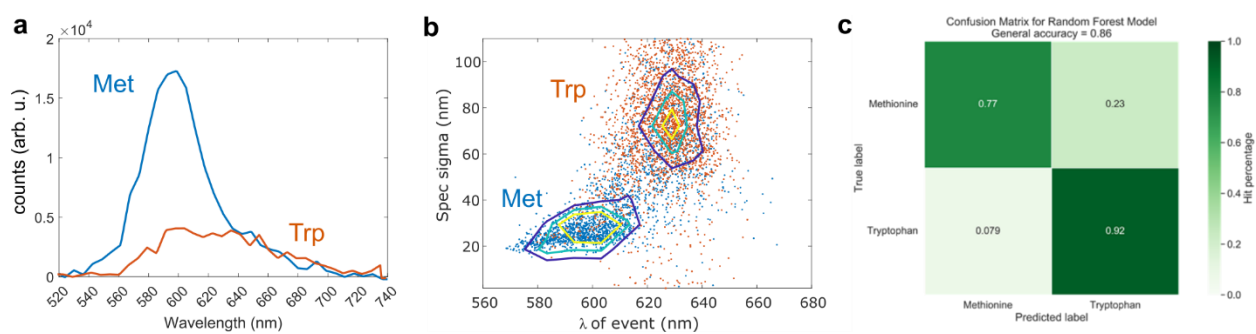

Figure S14. Manual identification of amino acids based on two features. **a**, Typical spectrum of methionine (Met) and tryptophan (Trp) activated single-defect emissions. The most probable emission spectrum was selected based on the emission histogram of each amino acids. **b**, Scatter plot of mean wavelength of events and spec sigma (spectral width by Gaussian fitting) for events activated by Met and Trp, respectively. For visualization purpose, 3000 events were randomly selected from each dataset. The contour lines indicate the number of events: 200 (yellow), 150 (blue), 100 (purple). **d**, A trade-off plot. **e**, Confusion matrix from Random Forest model shows general accuracy of 86% for Met and Trp.

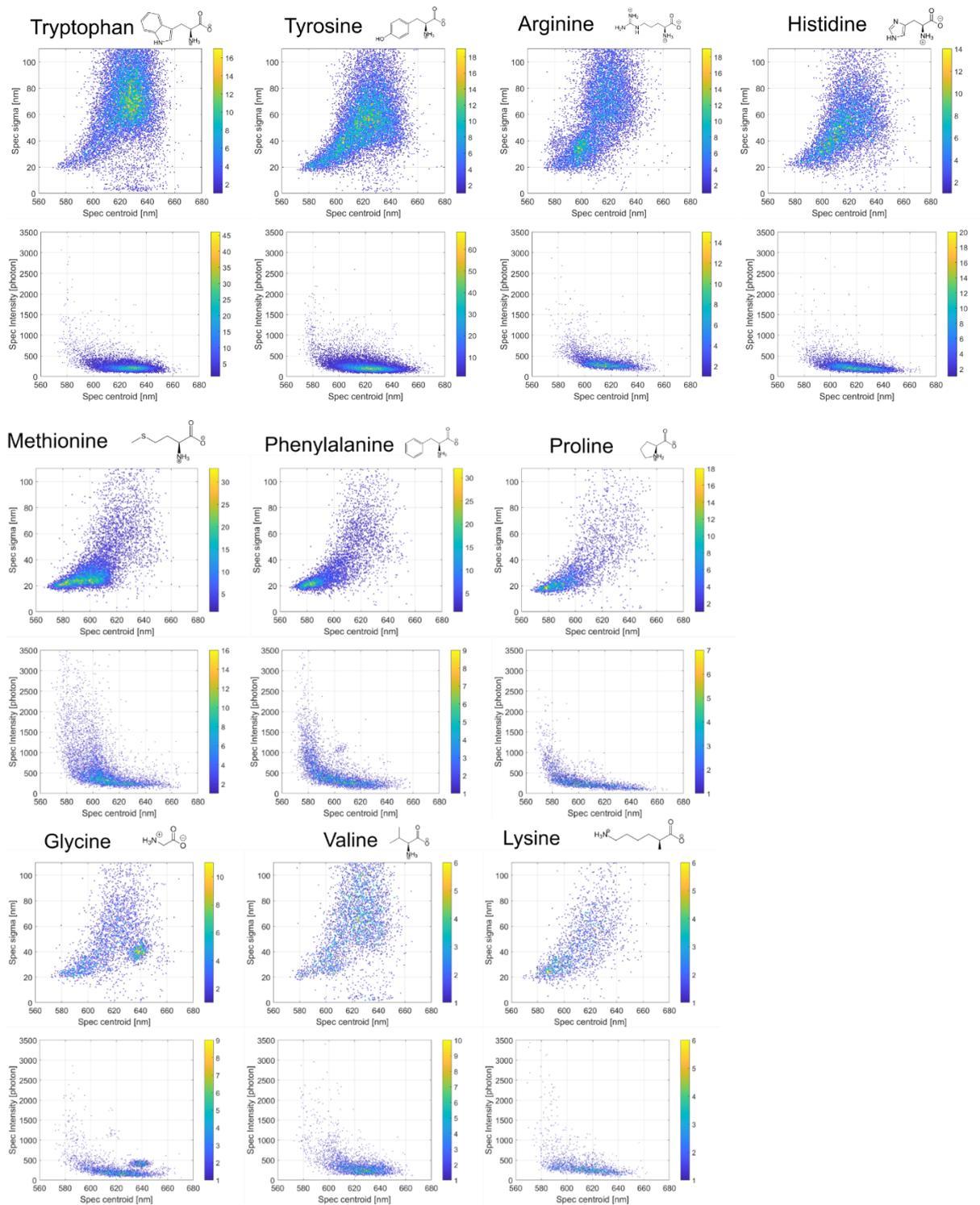

Figure. S15 Scatter plots of spec centroid versus spec sigma and spec centroid versus spec photons for 10 amino acids show the general correlation between spec photons, spec sigma and spec centroid.